\documentclass[aps,twocolumn,amsmath,amssymb,longbibliography,aps,prb]{revtex4-1}
\usepackage{graphicx,bm}
\usepackage{appendix}
\usepackage{color}
\usepackage{mathtools}
\usepackage{float}

\begin{document}

\title{Justification for zeta function regularization}
\author{F. R. Pratama.$^{1}$}
\email{pratama@flex.phys.tohoku.ac.jp}
\author{M.~Shoufie Ukhtary$^{1,2}$} 
\author{Riichiro Saito$^{1}$}
\affiliation{$^{1}$Department of Physics, Tohoku University, Sendai
  980-8578, Japan\\$^{2}$Research Center for Physics, Indonesian Institute of Sciences (LIPI), Tangerang Selatan 15314, Indonesia}

\begin{abstract}
    Using the fact that a finite sum of power series are given by the difference between two zeta functions, we justify the usage of the zeta function with a negative variable in physical problems to avoid the divergence of the infinite sum. We will show that in the case of magnetization of graphene, the zeta function with negative variable arises as a result of cut-off energy between two consecutive Landau levels. Furthermore, similar justification can be applied to the case of zero temperature Casimir force in parallel-plate geometry.
    \end{abstract}
\maketitle

The Riemann zeta function~\cite{riemann1859,elizalde12,elizalde21}, 
$\zeta(x)$, gives a non-physical value for $x\leqslant 0$. 
For example, $\zeta(-1)$ is given by
\begin{equation}
    \zeta(-1) = 1+2+3+4+\cdots = -\frac{1}{12}
    \label{eq:zeta}
\end{equation}
in which infinite sum of positive integers gives 
a negative fraction.
For a diverging summation that appears in physics,
if we adopt zeta function like 
Eq.\,(\ref{eq:zeta}), we {\it can} solve the physics problem, 
which is called zeta function regularization.
We usually hesitate to adopt the zeta function regularization. 
However, the non-physical values of zeta function, 
which can be understood mathematically by 
analytical continuation in the complex plane~\cite{riemann1859}, or by introducing a slowly decaying function in the diverging summation~\cite{tao13}, works surprisingly well for 
explaining the experimentally observed results. 
Thus we need justification of 
zeta function regularization.

One of the well-known examples of the application of zeta function regularization in physics is the Casimir force~\cite{casimir48,lamoreaux97,klimchitskaya09,rodriguez11,sushkov11}, which is observed in the cavity between two metallic plates~\cite{bressi02,decca07}. 
The Casimir force between two perfectly conducting plates at zero temperature ($T=0~\mathrm{K}$) is given by zeta function~\cite{hawking77,escobar20} to regularize infinite summation of zero-point energy~\cite{dalvit11}, in which a slowly decaying function is introduced to avoid divergence in the summation \cite{casimir48,hawking77,svaiter93,escobar20}. Another example of zeta function
regularization is magnetization of graphene\cite{ghosal07,pratama21-zeta}, where the summation of the Landau levels (LLs) is diverging.
Ghosal~\cite{ghosal07} calculated magnetization of graphene at the zero temperature by adopting zeta function regularization to avoid the divergence of thermodynamic potential. 
Recently, we also have adopted the zeta function regularization 
for calculating magnetization of the Dirac fermion as a function of magnetic field, temperature, and energy band gap~\cite{pratama21-zeta},
which reproduces the observed experimental results~\cite{li15}.
However, the zeta function regularization does not physically
justify the reason why we can use the zeta function except for the fact that the calculated results reproduces the 
experimental results. 

In this Letter, using the fact that a finite sum of power series can be represented by a difference between two zeta functions, we justify the zeta function regularization. The zeta functions appears only in the finite sum in the discussion. 
The present treatment can be generally used for any physics 
that has a divergence in the summation. 

Magnetization of undoped graphene, $M$ is given by the derivative of thermodynamic potential per unit area 
as follows:
\begin{equation}
    M(B) = -\frac{\partial\Delta\Omega(B)}{\partial B},
    \label{eq:M}
\end{equation}
where $\Delta\Omega (B)$ is the difference of thermodynamic potentials at a finite magnetic field $B$, $\Omega(B)$, 
from that for $B=0$, denoted by $\Omega^{(0)} $, as follows: 
\begin{equation}
    \Delta\Omega(B)\equiv \Omega(B)-\Omega^{(0)}.
    \label{eq:DO}
\end{equation}
Since $\Omega^{(0)}$ is not a function of $B$, 
we do not need to differentiate $\Omega^{(0)}$ by $B$ in Eq.\,(\ref{eq:M}). 
If we adopt the divergent summation in $\Omega(B)$ as a zeta function 
like Eq.\,(\ref{eq:zeta}), we obtain a finite value for $\Omega(B)$,
which corresponds to the zeta function regularization.
The role of $\Omega^{(0)}$ is to avoid the divergence of $\Delta\Omega(B)$
as a reference of energy.
It is clear from Eq.~(\ref{eq:DO}) that $\Delta\Omega(0)=0$. 
We will show 
that $\Delta\Omega(B)$ does not diverge for any value of $B$, 
though $\Omega(B)$ and $\Omega^{(0)}$ diverge. 

The LLs for graphene
within the approximation of the linear energy dispersion is given by\cite{pratama21-zeta}
\begin{equation}
    \epsilon_n = \mathrm{sgn}(n)\sqrt{2|n|} \frac{\hbar v_F}{\ell_B}\equiv  \mathrm{sgn}(n)\sqrt{|n|}\mathcal{E}(B),
    \label{eq:LL}
\end{equation}
where $\beta=1/(k_B T)$ ($k_B$ is the Boltzmann constant),
$\ell_B \equiv \sqrt{\hbar/(eB)}$ is the magnetic length, and 
$\mathcal{E}\equiv \sqrt{2}\hbar v_F/\ell_B$ is the  
$n=1$ LL energy. Let us first consider the case of strong field 
and low temperature ($\mathcal{E}\gg k_B T$) in undoped graphene.
Here, we only need to 
consider $\epsilon_n$ with $n\leqslant 0$ for $\Omega(B)$  as follows:
\begin{equation}
\begin{split}
    \Omega(B) &=  -\frac{4}{\beta}\frac{eB}{2h}\mathrm{ln}~2+\frac{4eB}{h} \sum_{n=-\infty}^{-1}\epsilon_n \\
    &\equiv \Omega_S (B)- C_B \sum_{n=1}^{\infty}\sqrt{n},
   \label{eq:omegaB}
   \end{split}
   \end{equation}
where $C_B$ is defined by $C_B = 2/(\pi {\ell_B}^2)\mathcal{E}(B)$ and has the unit of energy density [$\mathrm{J/m^2}$]. $\Omega_S $ is the contribution of the zeroth LL to $\Omega(B)$. $\Omega_S$ is associated with with entropy~\cite{pratama21-zeta} because of the linear dependence to $T$. For $\Omega^{(0)}$, we need to integrate states of electron at the valence band as follows: 
  \begin{equation}  
    \Omega^{(0)} = \int_{-\infty}^{0} \epsilon D(\epsilon) d\epsilon,
    \label{eq:omega0}
\end{equation}
where $\epsilon=\hbar v_F k$, and $D(\epsilon) = (2|\epsilon|/\pi)/(\hbar v_F)^2$ are the energy dispersion and the density of states per unit area of graphene
in the absence of the magnetic field, respectively. 

\begin{figure}[t]
\begin{center}
\includegraphics[width=80mm]{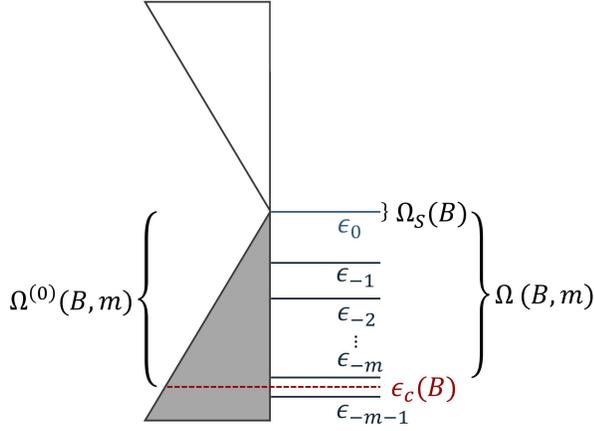}
\caption{ 
Schematic definitions of $\Omega^{(0)}(B,m)$, $\Omega(B,m)$, and $\Omega_S$ for $\mathcal{E}\gg k_B T$, where the cut-off index $c=-(m+1/2)$ for $\Omega^{(0)}(B,m)$. In the case of $\mathcal{E}\ll k_B T$, we also include electronic states at the conduction bands, where the limits of integration for $\Omega^{(0)}(B,m)$ is $\epsilon_{-m-x}\leqslant \epsilon \leqslant \epsilon_{m+x}$. Similarly, $\Omega(B,m)$ includes LLs ${-m}\leqslant n \leqslant {m}$. }\label{fig:lan}
\end{center}
\end{figure}

Since $\Delta\Omega(B)$ has one of mathematically indeterminate form such as $\Delta\Omega(B)= \infty - \infty$, we can not say that $\Delta\Omega(B)$ 
gives a finite value. 
In order to avoid the divergence of $\Omega(B)$ and $\Omega^{(0)}$, let us consider the energy cut-off for $\Omega(B)$ up to the 
$m$-th LL. The finite number of summation on 
$n=1$ to $m$ in the second term of Eq.~(\ref{eq:omegaB}) is given by the zeta functions as follows:
\begin{equation}
\begin{split}
 \Omega(B,m)&\equiv \Omega_S(B) -C_B\sum_{n=1}^{m}\sqrt{n} \\
 &=\Omega_S(B) - C_B\Big\{
 \zeta\Big(-\frac{1}{2}\Big) - \zeta\Big(-\frac{1}{2}, m+ 1\Big) \Big\}
  \label{eq:zeta1}
  \end{split}
\end{equation}
where $\zeta(s)$ and $\zeta(s, a)$ denote Riemann's and Hurwitz's zeta functions, respectively.\cite{olver-10}. The second term of Eq.\,(\ref{eq:zeta1}) can be checked to be correct for any integer values of $m$ by numerical calculation.
Practically, it is sufficient to check up to $m=10,000$.

For $m\gg 1$, we adopted the asymptotic expansion of $\zeta(-1/2, m+1)$ as follows~\cite{olver-10}: 
\begin{align}
    \zeta\Big(-\frac{1}{2},m+1 \Big) = -\frac{2}{3}m^{\frac{3}{2}} -\frac{1}{2}m^{\frac{1}{2}}-\frac{1}{24}m^{-\frac{1}{2}} +\mathcal{O}(m^{-\frac{3}{2}})
    \label{eq:zethm}
\end{align}
Further, we take an integration of Eq. (\ref{eq:omega0}) with an energy cut-off $\epsilon_c$ for $\Omega^{(0)}$, where $c\equiv -(m+x)$ as illustrated by Fig.~\ref{fig:lan}. When 
we choose $x=1/2$, the finite integral of $\Omega^{(0)}$ from $\epsilon_c$ to 0 becomes $B$ dependent because of the $B$ dependece of $\epsilon_c$ as follows, 
\begin{equation}
 \Omega^{(0)}(B,m) \equiv \int_{\epsilon_c}^{0} \epsilon D(\epsilon) d\epsilon
      = -\frac{2}{3}C_B \left(m + \frac{1}{2}\right)^{3/2}.
       \label{eq:omega0f}\\
\end{equation}
The factor $(m+1/2)^{3/2}$ can be expanded by 
 \begin{equation}
     \Big ( m +\frac{1}{2} \Big)^{3/2} = m^{3/2} +\frac{3}{4} m^{1/2} +\frac{3}{32} m^{-1/2} +\mathcal{O} (m^{-3/2})
     \label{eq:binomial}
 \end{equation}
Using Eqs. \,(\ref{eq:zeta1}) to
(\ref{eq:binomial}), we obtain 
$\Delta\Omega(B,m)\equiv \Omega(B,m)-\Omega^{(0)}(B,m)  $ as follows:
\begin{align}
    \Delta\Omega(B,m) =& \Omega_S(B)-C_B\Bigg[ \zeta\Big(-\frac{1}{2} \Big) -\frac{1}{48} m^{-1/2} \nonumber\\
    &+\mathcal{O} (m^{-3/2}) \Bigg].
    \label{eq:domegam}
\end{align}
Since the two terms that are proportional to $m^{3/2}$ and $m^{1/2}$
are cancelled to each other, the second term of $\Delta\Omega(B,m)$
are converged to a finite value of $-C_B\zeta(-1/2)$ in the limit of $m\to \infty$. 
From Eq.~(\ref{eq:domegam}), we obtain that $M$ of graphene for $\mathcal{E}\gg k_B T$ is given by 
$M(B) = \mathcal{C}_1 T + \mathcal{C}_2\sqrt{B}$, where $\mathcal{C}_1$ and $\mathcal{C}_2$ are 
constants~\cite{pratama21-zeta,li15}. 

It is important to note that we do not use zeta function regularization in Eq.\,(\ref{eq:domegam}). 
The value of zeta function $\zeta(-1/2)$ appears in the finite summation in Eq.\,(\ref{eq:zeta1}). The role of $\Omega^{(0)}(B,m)$ is to cancel the terms proportional to 
$m^{3/2}$ and $m^{1/2}$ as a reference of energy. 
 Thus we can justify the zeta function regularization for $\Omega(B)$ in which we do not need to consider $\Omega^{(0)}$ in Eq.\,(\ref{eq:M}).
It should be mentioned that the selection of $x=1/2$ in 
Eq.\,(\ref{eq:omega0f}) is essential to cancel the $m^{3/2}$ and $m^{1/2}$ terms. Generally, $x$ depends on $m$, which we discuss next.

Now, let us consider the case of weak-$B$ and high-$T$ limit for $M$ ($\mathcal{E}\ll k_B T$). In this limit, we need to consider the LLs from both valence and conduction bands. The $\Omega(B,m)$ and $\Omega(0,m)$ for $\mathcal{E}\ll k_B T$ are given by [See derivation in Section S1 B of the supplemental material]
\begin{equation}
\begin{split}
    \Omega(B,m) 
   & = 2\Omega_S^{(\mu)} + \sum_{\ell=0}^{\infty}C_\ell\left[\zeta (-\ell) - \zeta(-\ell, m+1)  \right]
   \end{split}
   \label{eq:omegaB-hT}
\end{equation}
and
\begin{equation}
\begin{split}
    \Omega^{(0)}(B,m)
    &=\sum_{\ell=0}^{\infty}C_\ell\frac{(m+x)^{\ell+1}}{\ell +1}
    \end{split}
    \label{eq:omega0-hT}
\end{equation}
where $\Omega_S^{(\mu)}=\ln{[1+\exp{(\beta\mu)}]}/(\pi{\ell_B}^2\beta)$ and $C_\ell$ is defined by
\begin{align}
    C_\ell=\frac{4}{\pi{\ell_B}^2\beta}\frac{(\beta\mathcal{E})^{2\ell}}{(2\ell)!}\mathrm{Li}_{1-2\ell}(-e^{\beta\mu}),
\end{align}
where $\mathrm{Li}_s(x)$ is polylogarithmic function. Therefore, the $\Delta\Omega(B,m)$ is given by a sum on $\ell$,
\begin{align}
    \Delta\Omega(B,m) = &~2\Omega_S^{(\mu)} + \sum_{\ell = 0}^{\infty}  C_\ell\Big[\zeta (-\ell)\nonumber \\
    &- \zeta(-\ell, m+1)- \frac{(m+x)^{\ell+1}}{\ell +1}\Big]
    \label{eq:DOH}.
\end{align}
Similar to Eq.\,(\ref{eq:zethm}), we expand $\zeta(-\ell, m + 1 )$
in terms of $m$ as follows~\cite{olver-10}: 
\begin{align}
    \zeta(-\ell,m+1) \sim &-\frac{m^{\ell+1}}{\ell+1} -\frac{1}{2}m^{\ell}\nonumber\\
   &+ \sum_{k=1}^{\infty}  \begin{pmatrix} -\ell+ 2k-2 \\ 2k-1 \end{pmatrix} \frac{\mathcal{B}_{2k}}{2k}m^{\ell + 1-2k},
    \label{eq:zeta-asymp}
\end{align}
where $\mathcal{B}_{2k}$ is the Bernoulli number. In order to cancel the terms $m^{\alpha}$ for $\alpha\geqslant 0$ while keeping $\zeta(-\ell)$, we obtain $x$ as a function of $m$ in the finite region of $0<x(m)<1$ that satisfies the following equation for each $\ell$, ($\ell = 0, 1, 2, \ldots)$,
\begin{align}
    \sum_{j=0}^{\ell} \begin{pmatrix} \ell+ 1 \\ j \end{pmatrix} \frac{m^j}{\ell + 1} x ^{\ell + 1 -j} -\frac{1}{2} m^{\ell} \nonumber&\\
    + \sum_{k=1}^{\lfloor\frac{\ell+1}{2}\rfloor}\begin{pmatrix} -\ell+ 2k-2 \\ 2k-1 \end{pmatrix}  \frac{\mathcal{B}_{2k}}{2k}m^{\ell + 1-2k} &= 0.
    \label{eq:polynomial}
\end{align}
If there exists a common  solution of $x(m)$ for all $\ell$'s in Eq.\,(\ref{eq:polynomial}), the zeta function regularization would be justified, which is not a trivial problem.
As shown in Supplemental materials, the solutions of $x(m)$ for each $\ell$
are not the same. However, we always find any $x(m)$ in the region of $0<x(m)<1$ for all $\ell$'s and if we take a limit of $m\rightarrow\infty$,
the solutions of $x(m)$ for all $\ell$'s are converging as follows
\begin{align}
    x(m)\rightarrow\frac{1}{2}~~~\mathrm{for~all}~\ell,~~m\rightarrow\infty,
    \label{eq:solution}
\end{align}
[See Eqs. (S7) - (S16) of the supplemental material]. The meaning of  Eq.~(\ref{eq:solution}) is that we can have the common $x$ in the limit of $m\to\infty$. By using Eq.~(\ref{eq:DOH}), magnetization of graphene for $\mathcal{E}\ll k_B T$ is given by a linear response $M(B)\propto -(B/T)\mathrm{sech}^2(\beta\mu/2)$ whose result~\cite{pratama21-zeta} 
is consistent with the formula of orbital susceptibility\cite{mcclure56}.

\begin{figure}[t]
\begin{center}
\includegraphics[width=60mm]{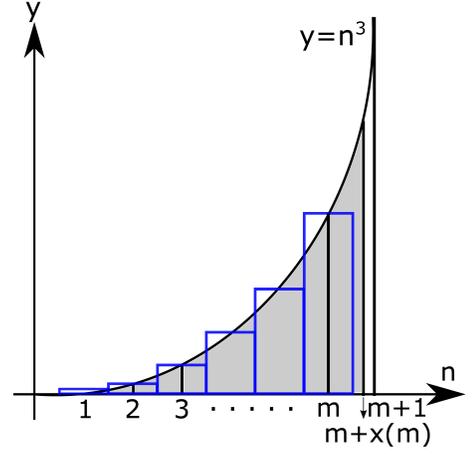}
\caption{ 
The illustration for $V(d)$ and $V^{(0)}$ given in Eq. (\ref{eq:delv}). The summation for $V(d)$ is taken up from $n=0$ to $m$, while the integration for $V^{(0)}$ is taken from $n=0$ to $m+x$, which is shown by the shaded region, to cancel the divergent terms in $V(d)$. }\label{fig:cas}
\end{center}
\end{figure}

Now let us discuss the justification for the case of 
the Casimir force.
The Casimir force is given by the derivative of the
summation of zero-point energy of electro-magnetic wave 
between two metallic plates separated by a distance $d$, 
\begin{equation}
    F_\textrm{C}=- \frac{\partial\Delta V(d)}{\partial d},
    \label{eq:cas0}
\end{equation}
 Similar to Eq. (\ref{eq:M}), the summation of zero-point energy  is given by $\Delta V(d)\equiv V(d)-V^{(0)}$, where $V(d)$ is the summation of zero-point energy when $d$ is sufficiently small compared with the one edge of the plates, $L$. $V^{(0)}$ is the reference energy of the vacuum.
 Similar to $\Omega^{(0)}$, the role of $V^{(0)}$ is to avoid the divergence for obtaining $F_\textrm{C}$. The $\Delta V(d)$ is expressed by [See Section S2 of the supplemental material for more detail derivation],
 \begin{align}
    \Delta V(d)&=V(d)-V^{(0)}\nonumber\\
    &=-\frac{C}{d^3}\left(\sum\limits_{n=0}^{m}n^3-\int\limits_{0}^{m+x}dn~ n^3\right), 
    \label{eq:delv}
 \end{align}
 where constant $C$ is given by,
 \begin{align}
 C=\frac{L^2\pi^2\hbar c}{6}.
 \end{align}
In Fig. \ref{fig:cas}, we illustrate for the summation in $V(d)$ and the integration in $V^{(0)}$, which is given by the shaded region.
Similar to the cases of magnetization, we use the cut-off index 
$m$ in the summation and integration of Eq. (\ref{eq:delv}). The 
$x$ in the cut-off for integration in Eq. (\ref{eq:delv}) is 
determined so that the $\Delta V$ does not diverge for large 
$m$. 
Similar with the case of magnetization, we can consider
the finite summation of 
Eq.\,(\ref{eq:delv}) as the subtraction of the two zeta functions as follows:
\begin{align}
     \sum\limits_{n=0}^{m}n^3 &= \zeta(-3) - \zeta(-3,m+1)\nonumber\\
     &=\zeta(-3)+\frac{m^4}{4}+ \frac{m^3}{2}+\frac{m^2}{2},
    \label{eq:casf}
\end{align}
where Eq. (\ref{eq:casf}) is obtained when we use the fact that $m\rightarrow\infty$ in evaluating the $\zeta(-3,m+1)$.  However, the divergent terms in $V(d)$ and $V^{(0)}$ are  cancelled to each other when we select $x=-m+m\sqrt{1+1/m}\approx 1/2+\mathcal{O}(m^{-1})$. In order to show how $\Delta V(d)$ is converged in the limit of $m\to\infty$, the $x$ is expanded for $1/m\ll 1$ up to fourth order [See Eq. (S23) of the supplemental material].
Finally we get $\Delta V(d)$ as follows:
\begin{align}
\Delta V(d)=-\frac{C}{d^3}\left[\zeta(-3)+\frac{7}{256m}+\mathcal{O}\left({m^{-2}}\right)\right],
\label{eq:delvd3}
\end{align}
where the terms in the bracket converge to $\zeta(-3)$ for increasing $m$.
Therefore, the $F_\textrm{C}$ is obtained from Eq. (\ref{eq:cas0}) as follows:
\begin{align}
    F_\textrm{C}=-\frac{3C}{d^4}\zeta(-3),
\end{align}
where $\zeta(-3)=1/120$. Thus, the two values of zeta functions
in Eq.\,(\ref{eq:casf}) has a physical meaning of the force. Once again, we do not use 
the zeta function regularization for obtaining the Casimir force, 
instead we consider a finite summation on $n^3$ in Eq.\,(\ref{eq:casf}).  

It is noted that the energies are measured from reference energies in the both cases of magnetization $\Delta \Omega$ and Casimir effect $\Delta V$. 
The reference energies are expressed by integration. 
In the Section S3 of the supplemental material, we also justify the zeta-function regularization when we use the definition of differential coefficient at a finite $B$
\begin{equation}
    M(B) =- \left. \frac{\partial \Omega(B)}{\partial B}\right|_{B} =- \lim_{\delta B \to 0}
    \frac{\Omega(B+\delta B)-\Omega(B)}{\delta B}.
    \label{eq:dcoeff}
\end{equation}
We can show that the $\Delta \Omega(B)$ (or $\Delta V(d)$) also converge to the corresponding zeta function by similar treatment.

In conclusion, zeta function regularization is justified by using the finite summation of energy up to finite discrete number $m$ that is subtracted by the reference energy. The zeta function appears as the remaining term when we subtract two diverging energies. The present treatment can be generally applied for any divergent summation in physics,
which should be useful for understanding the zeta function regularization.

FRP acknowledges MEXT scholarship. MSU and RS acknowledge JSPS KAKENHI Grant Number JP18H01810.

\bibliographystyle{apsrev4-1}

%





\end{document}


\preprint{APS/123-QED}

\title{Supplemental material: Justification for zeta function regularization }


	
\date{\today}

\maketitle


\section{Derivation of thermodynamic potential of graphene}

\subsection{Strong field and low temperature limit [$\mathcal{E}(B)\gg k_B T $]}

In  $\mathcal{E}(B)\gg k_B T $ limit, the spacing of the Landau levels (LLs) is much larger than $k_B T$. In the case of undoped graphene, therefore, we only need the to consider of $n\leqslant 0$ in the expression of $\Omega(B)$ as follows:
\begin{align}
    \Omega(B,m) = -\frac{1}{2}\frac{4}{\beta}\frac{eB}{h}\mathrm{ln}2-\frac{4}{\beta}\frac{eB}{h}\sum_{n=-m}^{-1}\mathrm{ln}[1+e^{-\beta\epsilon_n}] ,
    \label{eq:OB}
\end{align}
where $\beta = 1/(k_B T)$ and $eB/h=1/(2\pi{\ell_B}^2)$ is the Landau degeneracy (LD). The pre-factor $4$ accounts for the spin and valley degeneracies. It is noted that since the zeroth LL is half-occupied, the first term is multiplied by factor $1/2$. We use the fact that $-\beta\epsilon_n = \beta\mathcal{E}\sqrt{n}\gg 1$ for $n\geqslant 1$ to neglect the factor $1$ inside the logarithmic function in the second term of Eq. (\ref{eq:OB}), thus we get Eq.~(8) in the main text. 

The finite summation in Eq.~(8) of the main text is given in term of zeta functions. The Hurzwitz zeta function $\zeta(-\ell,m+1)$ in Eq.~(8) is expanded as a function of $m$ by the following expansion formula~\cite{olver-10},
\begin{align}
    \zeta(-\ell,m+1) = -\frac{m^{\ell+1}}{\ell+1} -\frac{1}{2}m^{\ell}+ \sum_{k=1}^{\infty}  \begin{pmatrix} -\ell+ 2k-2 \\ 2k-1 \end{pmatrix} \frac{\mathcal{B}_{2k}}{2k}m^{\ell + 1-2k},
    \label{eq:zeta-asymp}
\end{align}
where $\mathcal{B}_{2k}$ is the Bernoulli number. By considering only $k=1$ for the summation on $k$ in Eq. (\ref{eq:zeta-asymp}), we obtain Eq. (8) of the main text.

\subsection{Weak field and high temperature limit ($\mathcal{E}(B)\ll k_B T $)}

In  $\mathcal{E}(B)\ll k_B T $ limit, we need to consider the LLs in the range $-m \leqslant n \leqslant m$, because the spacing of LLs is much smaller than $k_B T$ and thus electrons can be thermally excited from the valence to conduction bands. Then, $\Omega(B;m)$ is given by
    \begin{align}
    \Omega(B,m) &= -\frac{4}{\beta}\frac{eB}{h}\sum_{n={-m}}^{m}\mathrm{ln}[1+e^{-\beta(\epsilon_n-\mu)}]\nonumber\\
   &=-\frac{2}{\pi{\ell_B}^2\beta}\mathrm{ln}[1+e^{\beta\mu}]  -\frac{2}{\pi{\ell_B}^2\beta}\Bigg[  \sum_{n=-m}^{-1} + \sum_{n=1}^{m} \Bigg]\sum_{k=1}^{\infty}\frac{(-1)^{k-1}}{k}e^{\beta\mu k}\sum_{\ell=0}^{\infty}\frac{(\pm\beta\mathcal{E}k)^{\ell}}{\ell!}|n|^{\ell/2}\nonumber\\
   & \equiv 2\Omega_S^{(\mu)} +  \frac{4}{\pi{\ell_B}^2\beta}\sum_{\ell=0}^{\infty}\sum_{k=1}^{\infty}\frac{(-e^{\beta\mu})^k}{k^{1-2\ell}}\frac{(\beta\mathcal{E})^{2\ell}}{(2\ell)!}\sum_{n=1}^{m} n^{\ell}\nonumber\\
   & = 2\Omega_S^{(\mu)} +  \frac{4}{\pi{\ell_B}^2\beta}\sum_{\ell=0}^{\infty}\mathrm{Li}_{1-2\ell}(-e^{\beta\mu})\frac{(\beta\mathcal{E})^{2\ell}}{(2\ell)!}[\zeta (-\ell) - \zeta(-\ell, m+1)  ]\nonumber\\
   & = 2\Omega_S^{(\mu)} + \sum_{\ell = 0}^{\infty} C_{\ell} [\zeta (-\ell) - \zeta(-\ell, m+1)  ]\nonumber\\
   & \equiv 2\Omega_S^{(\mu)} + \sum_{\ell=0}^{\infty}\Omega_{\ell}(m).
   \label{eq:omegaB-hT}
\end{align}
Here, $\Omega_{S}^{(\mu)}$ is defined by the first term in the second line and $\Omega_{S}^{(\mu)}=\Omega_S$ for the chemical potential $\mu=0$. 
In the second line of Eq.~(\ref{eq:omegaB-hT}), the first term is for $n=0$ and in the second term we expand the logarithmic and exponential functions for $n\neq 0$. It is noted that when we sum on $1 \le |n| \le m$, the terms for odd $\ell$'s cancel each other for the two summations of $n<0$ and $n>0$, while the terms for even $\ell$'s are doubled as shown in the third line. We switch the order of summations of $k$ and $\ell$ in order to express the $\mu$ dependence of $\Omega(B,m)$ by using polylogarithmic function in the fourth line, as follows:
\begin{align}
  \mathrm{Li}_s(x)=\sum_{k=1}^{\infty} \frac{x^{k}}{k^s}.
  \label{eq:polylog}
\end{align}
For simplicity, we defined the variable $C_{\ell}$ whose dimension is energy per unit area $\mathrm{[J/m^2]}$ as follows: 
\begin{align}
    C_{\ell} \equiv \frac{4}{\pi{\ell_B}^2\beta} \mathrm{Li}_{1-2\ell}(-e^{\beta\mu})\frac{(\beta\mathcal{E})^{2\ell}}{(2\ell)!}.
    \label{eq:clappend}
\end{align} 

In the calculation of $\Omega^{(0)}$, we need to consider states of electron in the whole Dirac cone. Let the energy cut-off at the valence band and conduction bands are given by $\epsilon_{-m-x}$  and $\epsilon_{m+x}$, we get
\begin{align}
    \Omega^{(0)}(B,m)&=-\frac{4}{\beta}\int_{\epsilon_{-m-x}}^{\epsilon_{m+x}}D(\epsilon) \mathrm{ln}[1+e^{-\beta(\epsilon-\mu)}]d\epsilon\nonumber\\
    &=\frac{2}{\pi\beta}\frac{1}{(\hbar v_F)^2}\sum_{\ell=0}^{\infty}\mathrm{Li}_{1-\ell}(-e^{\beta\mu})\frac{(-\beta)^\ell}{\ell!}\int_{\epsilon_{-m-x}}^{\epsilon_{m+x}} d\epsilon |\epsilon| \epsilon^\ell\nonumber\\
    &=\frac{2}{\pi\beta}\frac{1}{(\hbar v_F)^2}\sum_{\ell=0}^{\infty}\mathrm{Li}_{1-\ell}(-e^{\beta\mu})\frac{(-\beta)^\ell}{\ell!}\frac{\mathcal{E}^{\ell+2}}{\ell+2}\Bigg[ {(m+x)^{\frac{\ell+2}{2}}} - (-1)^{\ell+3}(m+x)^{\frac{\ell+2}{2}} \Bigg]\nonumber\\
    &=\frac{4}{\pi{\ell_B}^2\beta}\sum_{\ell=0}^{\infty}\mathrm{Li}_{1-2\ell}(-e^{\beta\mu})\frac{(\beta\mathcal{E})^{2\ell}}{(2\ell)!}\frac{(m+x)^{\ell+1}}{\ell+1}\nonumber\\
    &=\sum_{\ell=0}^{\infty}C_{\ell}\frac{(m+x)^{\ell+1}}{\ell+1}\nonumber\\
    &\equiv \sum_{\ell=0}^{\infty}\Omega_{\ell}^{(0)}.
    \label{eq:omega0-hT}
\end{align}
Therefore, the $\Delta\Omega(B,m)$ is given by an infinite series as follows:
\begin{align}
    \Delta\Omega(B,m) &= 2\Omega_S^{(\mu)} + \frac{4}{\pi{\ell_B}^2\beta}\sum_{\ell = 0}^{\infty}  \frac{(\beta\mathcal{E})^{2\ell}}{(2\ell)!}\mathrm{Li}_{1-2\ell}(-e^{\beta\mu})\Big[\zeta (-\ell)
    - \zeta(-\ell, m+1)- \frac{(m+x)^{\ell+1}}{\ell +1}\Big] \label{eq:delom}\\
    &\equiv 2\Omega_S^{(\mu)} + \sum_{\ell = 0}^{\infty} \Delta\Omega_{\ell}(B)
     \label{eq:delom2}
\end{align}
The zeta function $\zeta(-\ell, m+1)$ is expanded as a function of $m$ by Eq. (\ref{eq:zeta-asymp}). The second and third terms of Eq. (\ref{eq:delom}) gives polynomials of $m$. In order to cancels the terms $m^{\alpha}$ for $\alpha\geqslant 0$ while keeping $\zeta(-\ell)$, we look for a real solution of $0<x(m)<1$ that satisfies the following equation for each $\ell$,
\begin{align}
    \sum_{j=0}^{\ell} \begin{pmatrix} \ell+ 1 \\ j \end{pmatrix} \frac{m^j}{\ell + 1} x ^{\ell + 1 -j} -\frac{1}{2} m^{\ell} 
    + \sum_{k=1}^{\lfloor\frac{\ell+1}{2}\rfloor}\begin{pmatrix} -\ell+ 2k-2 \\ 2k-1 \end{pmatrix}  \frac{\mathcal{B}_{2k}}{2k}m^{\ell + 1-2k} = 0.
    \label{eq:polynomial}
\end{align}

Let us obtain analytical solutions $0<x(m)<1$ of Eq.~(\ref{eq:polynomial}) for several $\ell$'s. For $\ell=0$ the solution is exactly $x=1/2$ as follows:
\begin{align}
    x + m -m -\frac{1}{2} =0,
    \label{eq:l0}
\end{align}
By substituting $x=1/2$ into Eq.~(\ref{eq:delom2}), we get
\begin{align}
    \Delta\Omega_0(B) = \frac{4}{\pi{\ell_B}^2\beta}\mathrm{Li}_{1}(-e^{\beta\mu})\zeta(0) = -2 \Omega_{S}^{(\mu)},
    \label{eq:DO0}
\end{align}
where $\mathrm{Li}_1(z)=-\mathrm{ln}(1-z)$ and $\zeta(0)=-1/2$. Thus, we get $2\Omega_{S}^{(\mu)}+\Delta\Omega_0(B)=0$ in Eq. (\ref{eq:delom2}). Therefore, for $\mathcal{E}(B)\ll k_B T $, the $T$-dependence of $\Delta\Omega(B,m)$ is given as a power series of $\beta$ given in Eqs. (\ref{eq:omegaB-hT}) and (\ref{eq:omega0-hT}). 

For $\ell=1$, Eq.~(\ref{eq:polynomial}) is reduced to quadratic polynomial as follows:
\begin{align}
    \frac{1}{2}x^2 + mx -\Big( \frac{1}{2}m + \frac{1}{12} \Big) = 0.
    \label{eq:l1}
\end{align}
The solution for Eq.~(\ref{eq:l1}) for $m\gg 1$ and $\Delta\Omega_{1}$ are given by
\begin{align}
    x(m) = -m +\sqrt{m^2+m+\frac{1}{6}}\approx  \frac{1}{2} + \frac{1}{12 m} + \mathcal{O}(m^{-2}), 
    \label{eq:sol1}
\end{align}
and
\begin{align}
    \Delta\Omega_1(B) = -4\beta\frac{(v_F e B)^2}{\pi}\frac{e^{\beta\mu}}{(1+e^{\beta\mu})^2}\zeta(-1)  = \frac{(v_F e B)^2}{6\pi k_B T}\mathrm{sech}^2\Big( \frac{\mu}{2 k_B T} \Big),
    \label{eq:DO1}
\end{align}
respectively. In Eq.~(\ref{eq:DO1}) where we have used $\mathrm{Li}_{-1}(z) = z/(1-z) $ and $\zeta(-1) = -1/12$. From $\Delta\Omega_1$, we obtain that magnetization of graphene $M(B)\propto -B/T\mathrm{sech}^2(\beta\mu/2)$ for low field/high temperature.

For $\ell= 2$ and $\ell = 3$, we get a cubic and quartic polynomials from Eq.~(\ref{eq:polynomial}) as follows:
\begin{align}
    \frac{1}{3}x^3 + m x^2 + m^2 x -\Big( \frac{1}{2} m^2 + \frac{1}{6} m \Big ) = 0,
    \label{eq:l2}
\end{align}
and
\begin{align}
    \frac{1}{4}x^4 + m x^3 +\frac{3}{2} m^2 x^2  + m^3 x - \Big( \frac{1}{2}m^3 + \frac{1}{4}m^2 -\frac{1}{120} \Big) = 0,
    \label{eq:l3}
\end{align}
respectively. The solutions for Eqs.~(\ref{eq:l2}) and (\ref{eq:l3}) for $m\gg 1$  are given by
\begin{align}
    x(m) = -m + \Big(m^3+ \frac{3}{2}m^2 + \frac{1}{2}m\Big)^{1/3} \approx \frac{1}{2} + \frac{1}{6m} + \mathcal{O}(m^{-3}),
    \label{eq:sol2}
\end{align}
and 
\begin{align}
    x(m) = -m + \Big(m^4+ 2m^3 +m^2 -\frac{1}{30}\Big)^{1/4} \approx \frac{1}{2} + \frac{1}{4m} + \mathcal{O}(m^{-3}),
    \label{eq:sol2}
\end{align}
respectively. For $\ell\geqslant 4$, we have checked numerically that any solutions of Eq.~(\ref{eq:polynomial}) converges to $1/2$ for $m\gg 1$ (the Abel-Ruffini theorem states that the general solutions of polynomial $a_s x^s+ a_{s-1} x^{s-1} + \cdots =0 $ for $s\geqslant 5$ can not be expressed in radicals~\cite{rosen95}, i.e. by finite number of algebraic operations). It is noted that when we exclude constant term in the expression of the Hurwitz zeta function (i.e. by only including the terms $m^{\alpha}$ for $\alpha>0$ in Eq.~(\ref{eq:polynomial}), the solutions of $x(m)$ remains the same for large $m$. We adopt this treatment to derive the Casimir energy density in the next section, which is a special case for $\ell = 3$.

\section{The Derivation of Casimir effect}

The summation of zero-point energy  is given by $\Delta V\equiv V(d)-V^{(0)}$ as is given by Eq. (19) in the main text , where $V(d)$ is the summation of the zero-point energy when $d\ll L$ and $V^{(0)}$ is the reference point for measuring the energy, which is taken to be the zero-point energy when $d\gg L$. Similar to $\Omega^{(0)}$, $V^{(0)}$ is essential to avoid the divergence when we calculate $F_\textrm{C}$. The $V(d)$ is given as follows:
 \begin{align}
 V(d) &= \hbar c \frac{L^2}{\pi}\sum\limits_{n=0}^{\infty}\int\limits_{0}^{\infty}dk k\sqrt{\left(\frac{n\pi}{d}\right)^2+k^2}\label{eq:vd1}\\
 &\approx-\frac{L^2\hbar c}{6\pi}\left(\frac{\pi}{d}\right)^3\sum\limits_{n=0}^{m}n^3\equiv -\frac{C}{d^3}\sum\limits_{n=0}^{m}n^3\label{eq:vd2},
 \end{align}
where $k$ is the continuous wave vector in the direction parallel to the plate, while the $n$ is the index for quantized wave vector in the direction perpendicular to the plate since the $d\ll L$. We also define constant $C$ as follows:
\begin{align}
 C=\frac{L^2\pi^2\hbar c}{6}\label{eq:C}.
 \end{align}
The $V^{(0)}$ is obtained by replacing the summation on $n$ by integration, since the $d\gg L$. The $V^{(0)}$ is expressed as follows:
\begin{align}
  V^{(0)}&=\hbar c \frac{L^2}{\pi}\frac{d}{\pi}\int\limits_{0}^{\infty}dn\int\limits_{0}^{\infty}dk k\sqrt{\left(\frac{n\pi}{d}\right)^2+k^2}\label{eq:v01}\\ 
  &\approx-\frac{L^2\hbar c}{6\pi}\left(\frac{\pi}{d}\right)^3\int\limits_{0}^{m+x}dn~ n^3\equiv -\frac{C}{d^3}\int\limits_{0}^{m+x}dn~ n^3\label{eq:v02},
\end{align}
where we adopt the definition of $C$ given in Eq. (\ref{eq:C}). It is noted that, the energies for both $V(d)$ and $V^{(0)}$ are measured from the infinite energy coming from the evaluation of upper limit of the integration with respect to $n$.

The $V^{(0)}$ cancels exactly the divergent terms of $V(d)$ given in Eq. (21) of the main text when we choose $x=-m+\sqrt{m^2+m}$, since we have,
\begin{align}
   \int\limits_{0}^{\sqrt{m^2+m}}dn\, n^3=\frac{1}{4}\left(m^2+m\right)^2\nonumber
   =\frac{m^4}{4}+ \frac{m^3}{2}+\frac{m^2}{2}.
\end{align}
Therefore, the $\Delta V(d)$ is given as follows:
\begin{align}
\Delta V(d)=-\frac{C}{d^3}\zeta(-3),
\label{eq:delvd2}
\end{align}
which is finite and does not depend on $m$.  The $F_\textrm{C}$ is given as follows:
\begin{align}
    F_\textrm{C}=-\frac{3C}{d^4}\zeta(-3)=-\frac{L^2\hbar c}{240}\frac{\pi^2}{d^4},\label{eq:fcas}
\end{align}
where we use $\zeta(-3)=1/120$. To see how the convergence of $\Delta V(d)$ is reached, let us Taylor expand the $x=-m+m\sqrt{1+1/m}$ for $1/m\ll 1$. When we expand the $x$ up to fourth order of $1/m$, we obtain 
\begin{align}
    x\approx-m+m\left(1+\frac{1}{2m}-\frac{1}{8m^2}+\frac{1}{16m^3}-\frac{5}{128m^4}\right).\label{eq:xexpand}
\end{align}
The corresponding $\Delta V(d)$ is expressed as follows:
\begin{align}
\Delta V(d)=-\frac{C}{d^3}\left(\zeta(-3)+\frac{7}{256m}+\mathcal{O}\left(\frac{1}{m^2}\right)\right),
\label{eq:delvd3}
\end{align}
which converges to Eq. (\ref{eq:delvd2}) by taking a limit of $m\rightarrow \infty$, similar to the case of magnetization in Eq. (11) of the main text.

\section{Alternative derivation of the magnetization and the casimir effect}

Here, we will show an alternative derivation of magnetization and the Casimir effect by using the quantized reference point for measuring the energy. Let us start with the derivation of magnetization. In this case, the $\Omega^{(0)}(B,m')$ is quantized since the magnetic field is changed infinitesimally from the $\Omega(B,m)$ to $\Omega(B+\delta B,m')$. The $\Delta \Omega(B,m)$ is given by,
\begin{align}
   \Delta \Omega(B,m) &=  \Omega (B+\delta B,m')-\Omega(B,m) \nonumber\\
   &= -\frac{4}{\beta}\frac{e\delta B}{2h}\ln{2} -\frac{v_F}{\sqrt{\hbar\pi^2}}\left[\left(2e(B+\delta B)\right)^{\frac{3}{2}}\sum_{n=1}^{m'}\sqrt{n'}-\left(2eB\right)^{\frac{3}{2}}\sum_{n=1}^{m}\sqrt{n}\right]\nonumber\\
   &= -\frac{4}{\beta}\frac{e\delta B}{2h}\ln{2} -\frac{v_F}{\sqrt{\hbar\pi^2}}\left(2eB\right)^{\frac{3}{2}}\left[\left(1+\frac{3}{2}\frac{\delta B}{B}\right)\sum_{n=1}^{m'}\sqrt{n'}-\sum_{n=1}^{m}\sqrt{n}\right]\nonumber\\
   &\equiv-\frac{4}{\beta}\frac{e\delta B}{2h}\ln{2} -C_B\left[\left(1+\frac{3}{2}\frac{\delta B}{B}\right)\sum_{n=1}^{m'}\sqrt{n'}-\sum_{n=1}^{m}\sqrt{n}\right]\label{eq:delome},
\end{align}
where the reference point for measuring energy is not $\Omega^{(0)}$ but $\Omega(B+\delta B,m')$. We also define, 
\begin{align}
    C_B = \frac{v_F}{\sqrt{\hbar\pi^2}}\left(2eB\right)^{\frac{3}{2}}\equiv \frac{2}{\ell_B^2}\mathcal{E}(B).\label{eq:cb}
\end{align}
Let us determine the $m'$ as a function of $m$ by supposing the $m$ and $m'$ have the same LL, that is,
\begin{align}
    \sqrt{2\hbar v_F^2 eBm}=\sqrt{2\hbar v_F^2 e(B+\delta B)m'}.
\end{align}
We obtain the following relationship,
\begin{align}
    m' =\left(1-x\frac{\delta B}{B}\right)m,\label{eq:mprimmag}
\end{align}
where we add a parameter $x$ to cancel the diverging terms. By expressing the finite summation in terms of zeta functions, Eq. (\ref{eq:delome}) becomes,
\begin{align}
   \Delta \Omega(B,m) 
   = -\frac{4}{\beta}\frac{e\delta B}{2h}\ln{2} &-C_B\Bigg[\left(1+\frac{3}{2}\frac{\delta B}{B}\right)\left(\zeta(-1/2)+\frac{2}{3}m'^{\frac{3}{2}}+\frac{1}{2}m'^{\frac{1}{2}}\right)\nonumber\\
   &-\zeta(-1/2)+\frac{2}{3}m^{\frac{3}{2}}+\frac{1}{2}m^{\frac{1}{2}}+\mathcal{O}(m^{-\frac{1}{2}})\Bigg]\label{eq:delome2}.
\end{align}
By substituting Eq. (\ref{eq:mprimmag}) to Eq. (\ref{eq:delome2}) and by considering only terms linear to $\delta B$, we obtain the following equation,
\begin{align}
    \Delta \Omega(B,m) 
   = -\frac{4}{\beta}\frac{e\delta B}{2h}\ln{2} -C_B\frac{\delta B}{B}\Bigg[\frac{3}{2}\zeta(-1/2)-(x-1)m^{\frac{3}{2}}-\frac{1}{4}(x-3)m^{\frac{1}{2}}+\mathcal{O}(m^{-\frac{1}{2}})\Bigg]\label{eq:delome3},
\end{align}
which is diverging with increasing $m$. By choosing the $x$ to be,
\begin{align}
    x = \frac{4m+3}{4m+1},
\end{align}
all the diverging terms of $m$ vanish and the term in the bracket of Eq. (\ref{eq:delome3}) converges to $(3/2) \zeta(-1/2)$. The magnetization is obtained by substituting Eq. (\ref{eq:delome3}) to Eq. (24) in the main text.

Let us consider the Casimir effect. In the Eq. (19) of the main text, the summation of zero-point energy is measured from the reference point when the $d \gg L$. Here, we will show that the Casimir effect is also obtained when we consider the reference point with width of $d'\ll L$, that is, $d'=d+\delta d$. In this case, the $\Delta V(d)$ is expressed as follows:
\begin{align}
\Delta V(d)&=-\frac{L^2\hbar c}{6\pi}\left(\left(\frac{\pi}{d+\delta d}\right)^3\sum\limits_{n'=0}^{m'}n'^3-\left(\frac{\pi}{d}\right)^3\sum\limits_{n=0}^{m}n^3\right)
\label{eq:delvd1s}\\
&=-\frac{C}{d^3}\left(\left(1-3\frac{\delta d}{d}\right)\sum\limits_{n'=0}^{m'}n'^3-\sum\limits_{n=0}^{m}n^3\right)
\label{eq:delvd2s}
\end{align}
where the reference point is also quantized and $\delta d \ll d$.  Here, the constant $C$ is defined in Eq. (\ref{eq:C}). Let us consider the cut-off index $m'$ as a function of $m$. By supposing that the cut-off indices for the two energies have the same wave vector that is, 
\begin{align}
    \frac{m\pi}{d}=\frac{m'\pi}{d+\delta d},
\end{align}
we obtain the following relation,
\begin{align}
    m' = \left(1+x\frac{\delta d}{d}\right)m,\label{eq:mprime}
\end{align}
where we add the parameter $x$ to cancel the diverging terms. By expressing the finite summation in terms of zeta functions, Eq. (\ref{eq:delvd2s}) is expressed as follows:
\begin{align}
   \Delta V(d)&= -\frac{C}{d^3}\left[\left(1-3\frac{\delta d}{d}\right)\left(\zeta(-3)+\frac{m'^4}{4}+\frac{m'^3}{2}+\frac{m'^2}{4}\right)-\zeta(-3)+\frac{m^4}{4}+\frac{m^3}{2}+\frac{m^2}{4}\right]\label{eq:delvd3s}
\end{align}
By substituting the $m'$ with Eq. (\ref{eq:mprime}) and considering only term linear to $\delta d$, the $\Delta V(d)$ is written as follows:
\begin{align}
    \Delta V(d)= -C\frac{\delta d}{d^4}\left(-3\zeta(-3)+(4x-3)\frac{m^4}{4}+(x-1)\frac{m^3}{2}+(2x-3)\frac{m^2}{4}\right),\label{eq:delvd4s}
\end{align}
which is diverging with increasing $m$. By choosing the $x$ to be
\begin{align}
    x = \frac{3m^2+2m+3}{2(2m^2+m+1) },
\end{align}
all the diverging terms in Eq. (\ref{eq:delvd4s}) vanish and the terms in the bracket of Eq. (\ref{eq:delvd4s}) becomes $-3\zeta(-3)$. The $F_\textrm{C}$, which is defined by Eq. (18) in the main text, is obtained by dividing the Eq. (\ref{eq:delvd4s}) with $\delta d$ and multiplying with $-1$, which gives Eq. (\ref{eq:fcas}).

The two methods are equivalent by taking proper $x$ that is not the same for the two methods, even if we obtain magnetization at a finite $B$ or the Casimir force at a finite $d$. In the first method, we take derivative $\partial \Delta\Omega(B) / \partial B$ or $\partial \Delta V(d)/ \partial d$ and then we put the finite value of $B$ or $d$. In the second method, we adopt the definition of slope of the $\Omega (B)$ or $V(d)$, that is, 
\begin{align}
    \lim_{\delta B \to 0}
    \frac{\Omega(B+\delta B)-\Omega(B)}{\delta B}~~~~~~~\textrm{or}~~~~~~\lim_{\delta d \to 0}
    \frac{V(d+\delta d)-V(d)}{\delta d}.
\end{align}

\bibliographystyle{apsrev4-1}